\documentstyle[preprint,aps]{revtex}

\newcommand{\rgrad}{{\rm grad}}
\newcommand{\rdiv}{{\rm div}}
\newcommand{\bg}[1]{\mbox{\boldmath{$ #1 $}}}
\newcommand{\hdot}{\mbox{\huge .}}

\begin{document}

\title{Two-phase binary fluids and immiscible fluids described by an
order parameter}
\author{Morton E. Gurtin}
\address{Department of Mathematics, Carnegie Mellon University,
Pittsburgh, Pennsylvania 15213}
\author{Debra Polignone}
\address{Department of Mathematics, University of Tennessee, Knoxville,
Tennessee 37996}
\author{Jorge Vi\~nals}
\address{Supercomputer Computations Research Institute, Florida State
University, Tallahassee, Florida 32306, and Department of Chemical
Engineering, FAMU/FSU College of Engineering, Tallahassee, Florida 32310}

\date{\today}

\maketitle

\begin{abstract}

A unified framework for coupled
Navier-Stokes/Cahn-Hilliard equations is developed using, as a basis, a
balance law for microforces in conjunction with constitutive equations
consistent with a mechanical version of the second law. As a numerical
application of the theory, we consider the kinetics of coarsening for a
binary fluid in two space-dimensions.

\end{abstract}


\section{Introduction}

The Cahn-Hilliard equation \cite{re:cahn61}
\footnote{Notation. Tensors are linear
transformations of $R^{3}$ into $R^{3}$ and are denoted by upper-case boldface
letters. Vectors may be viewed as 3x1 column vectors and tensors as
3x3 matrices. {\bf 1}
denotes the unit tensor; ${\bf a} \otimes {\bf b}$, the tensor product of
vectors ${\bf a}$ and ${\bf b}$, is the tensor defined
by $({\bf a} \otimes {\bf b}) {\bf u} =({\bf b} \cdot {\bf u}){\bf a} $
for all vectors ${\bf u}$; ${\bf A}^{T}$ is the transpose of a tensor
${\bf A}$; tr ${\bf A}$ is the trace
of ${\bf A}$; the inner product of tensors ${\bf A}$ and ${\bf B}$ is
defined by ${\bf A} \cdot {\bf B} = {\rm tr} ({\bf A}^{T} {\bf B})$.

The divergence, gradient, and Laplacian of a field $\varphi = \varphi
({\bf x},t)$ are
denoted by $\rgrad \varphi$, ${\rm div} \varphi$, and $\Delta \varphi$.
For a vector field ${\bf u}({\bf x})$, $\rgrad {\bf u}({\bf x})$
is the tensor with components $\partial u_{i} / \partial x_{j}$
(i = row index, j = column index). The divergence of a tensor field
${\bf A}({\bf x})$ is the vector field with
components $\sum_{j} \partial A_{ij}/\partial x_{j}$ (i=row index).

The derivative of a function $f$ of a scalar variable (not time) is denoted
by a prime: $f'$. The partial
derivative of a function $\Phi(a,b,c,\ldots, d)$ (of n scalar, vector,
or tensor variables) with respect to b, say, is written $\partial_{b}
\Phi (a,b,c, \ldots , d)$.}
\begin{equation}
\varphi^{\hdot} = m \Delta \left[ f'(\varphi) - \alpha \Delta \varphi
\right]
\end{equation}
is central to materials science, as it characterizes important qualitative
features of two-phase systems. This equation is based on a free energy
\begin{equation}
\label{eq:fe}
\hat{\psi}(\varphi, \rgrad \varphi) = f(\varphi) + \frac{1}{2} \alpha |
\rgrad \varphi |^{2},
\end{equation}
with $f(\varphi)$ a double-well potential whose wells define
the phases, and leads to an interfacial layer within which the density
$\varphi$ suffers large variations.

The standard derivation of the Cahn-Hilliard equation begins with the
mass balance
\begin{equation}
\varphi^{\hdot} = - {\rm div} ~ {\bf h}
\end{equation}
and the constitutive equation
\begin{equation}
{\bf h} = - m ~ \rgrad \mu,
\end{equation}
which relates the mass flux ${\bf h}$ to the chemical potential $\mu$. The
presence of density gradients renders (\ref{eq:fe}) incompatible with the
classical definition of $\mu$ as the partial derivative of $\psi$ with respect
to $\varphi$; instead $\mu$ is defined as the variational derivative
$\mu = \delta \Psi / \delta \varphi$ of the total free energy
\begin{equation}
\Psi (\varphi) = \int_{B} \hat{\psi} (\varphi, \rgrad \varphi) dv,  ~~~
{\rm B ~ = ~ underlying ~ region ~ of ~ space}.
\end{equation}
Since
\begin{equation}
\delta \Psi / \delta \varphi = f'(\varphi) - \alpha \Delta \varphi,
\end{equation}
this yields the Cahn-Hilliard equation.

The major advances in nonlinear
continuum mechanics over the past thirty years are based on the separation
of balance laws (such as those for mass and force), which are general and
hold for large classes of materials, from constitutive equations (such as
those for elastic solids and viscous fluids), which delineate specific
classes of material behavior. In the derivation presented above there is no
such separation, and it is not clear whether there is an underlying
balance law that can form a basis for more general theories.

An alternative derivation of the Cahn-Hilliard equation \cite{re:gurtin95} is
based on a balance law for microforces \cite{re:fried93}, defined
operationally as forces whose working accompanies changes in $\varphi$. These
forces are described by a (vector) stress \bg{\xi}, which characterizes
forces transmitted across surfaces, and a (scalar) body force $\pi$, which
represents internal forces distributed over the volume of the material.
The basic hypothesis is the microforce balance
\begin{equation}
\label{eq:microforce}
\int_{\partial R} \bg{\xi} \cdot {\bf n} da + \int_{R} \pi dv = 0
\end{equation}
for each control volume $R$, with ${\bf n}$ the outward unit normal
to $\partial R$.

Here we study the isothermal motion of an incompressible binary
fluid, with one of the constituent densities serving as the order
parameter. A chief assumption is that the relative momenta and kinetic
energies of the constituents are negligible when computed relative to
the gross motion of the fluid. This allows us to base the theory on:
\begin{itemize}
\item balance of mass for the order parameter;
\item the microforce balance (\ref{eq:microforce});
\item balance of momentum for the macroscopic motion of the fluid;
\item a version of the second law in the form of a global
free-energy inequality in which the microscopic working \cite{re:fried93}
\begin{equation}
\int_{\partial R} \varphi^{\hdot} ~ \bg{\xi} \cdot {\bf n} ~ da
\end{equation}
joins the working of the macroscopic stresses; and
\item constitutive equations presumed compatible with this version of
the second law.
\end{itemize}

In standard theories of diffusion the chemical
potential is given, constitutively, as a function of the density, but
here we consider systems sufficiently far from equilibrium that a relation
of this type is no longer valid; instead we allow the chemical potential
and its gradient to join the stretching tensor, the density, and the
density gradient as independent constitutive variables. Interestingly,
the second law reduces the free energy to a function $\psi = \hat{\psi}
(\varphi, \rgrad \varphi)$ of at most $\varphi$ and $\rgrad \varphi $,
where $\rgrad \varphi$ here denotes the spatial (Eulerian) gradient.

A chief result of our theory is a constitutive equation
giving the macroscopic (Cauchy) stress as a classical Newtonian stress
plus a term
\begin{equation}
\alpha ~ \rgrad \varphi \otimes \rgrad \varphi,
\end{equation}
which represents capillarity. \footnote{A term of this type appears in
Korteweg's theory of gradient fluids (cf. Truesdell \& Noll
\cite{re:truesdell65})}
Our final results consist of the coupled Navier-Stokes/Cahn-Hilliard equations
\begin{eqnarray}
\label{eq:model}
\rho \left[ {\bf v}_{t} + (\rgrad {\bf v}) {\bf v} \right] & = & -
\rgrad p + \nu \Delta {\bf v} - \alpha ( \Delta \varphi) \rgrad
\varphi, ~~~ {\rm div} {\bf v} =0, \nonumber \\
\varphi_{t} + ( \rgrad \varphi) \cdot {\bf v} & = & m \Delta \left[
f'(\varphi)- \alpha \Delta \varphi \right].
\end{eqnarray}

We also develop equations
for the motion of immiscible fluids. Here our use of an order parameter
endows the interface with capillarity and yields a theory in which the
free-boundary need not be explicitly tracked; as this theory does not allow
for a phase change, the interface is transported with the material.

The model defined by (\ref{eq:model}),
which is referred to as \lq\lq Model H" in the
literature on critical phenomena \cite{re:hohenberg77}, was used
by Siggia, Halperin, \& Hohenberg \cite{re:siggia76} to study behavior at the
critical points of single and binary fluids. Later extensions of the model
were used to analyze the superfluid transition of $^{4}$He
\cite{re:pankert81} as well as the decay of critical fluctuations in simple
fluids \cite{re:onuki79} and polymer solutions \cite{re:helfand89}
under an imposed shear flow. A chief purpose of these studies was to
analyze the effect of reversible modes (i.e., those corresponding to
the coupling terms ($\rgrad \varphi) \cdot {\bf v}$ and
$\alpha (\Delta \varphi) \rgrad \varphi$) on the dynamical
critical behavior of otherwise purely dissipative models, with emphasis
placed on the behavior of hydrodynamic transport coefficients near the
critical point. More recently, (\ref{eq:model}) was used to study other
nonequilibrium phenomena below the critical point; for example, Kawasaki
\& Ohta \cite{re:kawasaki83} and Koga \& Kawasaki \cite{re:koga91}
use (\ref{eq:model}) to study hydrodynamic effects during spinodal
decomposition.

The applicability of
(\ref{eq:model}) to studies of the type discussed above is generally based, not
on first principles, but instead on the observation that a convective
term ($\rgrad \varphi) \cdot {\bf v}$ in the Cahn-Hilliard equation should
be accompanied by a term $\alpha (\Delta \varphi) \rgrad \varphi$
in the Navier-Stokes equation to ensure that
the asymptotic probability-distribution functions obtained as
$t \rightarrow \infty $
coincide with those given by standard equilibrium calculations. In view
of the phenomenological nature of this formalism, it seems natural to
ask whether there is a derivation of (\ref{eq:model}) within the framework of
nonlinear continuum mechanics. Here we provide such a derivation.

In addition, we present computational results appropriate to a range of
parameters in which sharp interfaces (large, localized changes in $\varphi$)
exist.

{}From a computational standpoint our approach is similar
to VOF (volume-of-fluid) methods \cite{re:hirt81}. There a color
function, which plays a role similar to that of $\varphi$, is passively
advected by the flow. More recently the VOF method has been extended to
include capillary effects \cite{re:unverdi92,re:haj-hariri94}
leading to a modified Navier-Stokes equation equivalent to
the first of (\ref{eq:model}). These models, however, do not endow the color
function with any physical significance, and dissipative relaxation as
in the equation for $\varphi$ in (\ref{eq:model})
is absent. In comparison, not only does the model based on (\ref{eq:model})
include capillary effects in the modified
Navier-Stokes equation, it also explicitly includes other dissipative
processes up to the length scale of the smeared interface; it can
therefore describe events such as break-up and coalescence, with the
proviso that the physical behavior at those length scales is determined
by the choice of free energy $\psi$.

\section{Basic Laws}

We consider the isothermal motion
of an incompressible binary fluid whose total density is constant. A
basic assumption of our theory is that {\em the momenta and kinetic energies
of the constituents are negligible when computed relative to the gross
motion of the fluid}.

\subsection{Kinematics}

We write $\rgrad$ and $\rdiv$
the spatial gradient and spatial divergence. Incompressibility then
requires that ${\bf v}$, the gross velocity of the fluid, satisfy
\begin{equation}
{\rm div} ~ {\bf v} = {\rm tr} ~ ( \rgrad  {\bf v} ) = 0,
\end{equation}
a constraint that renders
the stretching,
\begin{equation}
{\bf D} = \frac{1}{2} \left( \rgrad  {\bf v} + \rgrad
{\bf v}^{T} \right)
\end{equation}
traceless.

We will use both the spatial time derivative $(...)_{t}$ and
the material time derivative $(...)^{\hdot}$; these are related through
\begin{equation}
(...)^{\hdot} = (...)_{t} + \left[ \rgrad (...) \right] {\bf v}.
\end{equation}
We will also
use the following identity for scalar fields $\varphi$,
\begin{equation}
\label{eq:mathident}
\rgrad ( \varphi^{\hdot} ) = ( \rgrad \varphi)^{\hdot} + (\rgrad
{\bf v})^{T} \rgrad \varphi .
\end{equation}

\subsection{Balance of mass}

We write $\rho_{b}$ for the density and ${\bf h}_{b}$ for the mass flux
of constituent $b ~ (=1, 2)$, with ${\bf h}_{b}$ measured relative to the gross
motion of the fluid. Then
\begin{equation}
\label{eq:mcon}
\rho_{1} + \rho_{2} = \rho
\end{equation}
with $\rho$, the total density
of the fluid, constant; consistent with this constraint, we assume
that
\begin{equation}
\label{eq:pcon}
{\bf h}_{1} + {\bf h}_{2} = 0.
\end{equation}

Throughout the
paper $R$ denotes an arbitrary  control volume (fixed region of space),
with ${\bf n}$ the outward unit normal to $\partial R$. Given a field $\Phi$,
\begin{equation}
\left\{ \int_{R} \Phi dv \right\}^{\hdot} = \int_{R} \Phi^{\hdot} dv,
\end{equation}
where the dot signifies the time derivative following the material currently
in R, and where the integral and time derivative commute by virtue of
balance of mass, since the total density of the fluid is constant.

Balance of mass requires that
\begin{equation}
\label{eq:mbal}
\left\{ \int_{R} \rho_{b} dv \right\}^{\hdot} = - \int_{\partial R}
{\bf h}_{b} \cdot {\bf n} ~ da
\end{equation}
for each constituent $b$. Because of (\ref{eq:mcon}) and
(\ref{eq:pcon}), one of the two relations (\ref{eq:mbal}) is redundant;
we therefore let
\begin{equation}
\varphi = \rho_{1} ~~~ {\bf h} = {\bf h}_{1} ,
\end{equation}
and restrict attention to the local balance
\begin{equation}
\label{eq:phicon}
\varphi^{\hdot} = - {\rm div} {\bf h}
\end{equation}
for the first constituent.

\subsection{Balance of momentum}

We write ${\bf T}$ for the stress tensor associated with
the macroscopic motion of the fluid. Then, since we neglect the relative
momenta of the constituents, the balance laws for linear and angular
momentum have the standard form
\begin{equation}
\int_{\partial R} {\bf T} {\bf n} da = \left\{ \int_{R} \rho {\bf v} dv
\right\}^{\hdot} ~~ \int_{\partial R} {\bf x} \times {\bf T} {\bf n} da =
\left\{ \int_{R} ( {\bf x} \times \rho {\bf v} ) dv \right\}^{\hdot}
\end{equation}
for each $R$, or equivalently,
\begin{equation}
\label{eq:force}
{\rm div} {\bf T} = \rho {\bf v}^{\hdot} ~~~ {\bf T} = {\bf T}^{T}.
\end{equation}
It is convenient to introduce the extra
stress ${\bf S}$ and the pressure $p$ defined by
\begin{equation}
\label{eq:extras}
{\bf S} = {\bf T} + p {\bf 1} ~~~~ p = - \frac{1}{3} ({\rm tr} ~ {\bf T}) {\bf
1}.
\end{equation}
As a consequence of incompressibility, $p$ is indeterminate.

\subsection{Order parameter. Microforce balance}

We assume that the microscopic behavior
of the fluid, as manifested in the diffusion of its constituents,
is described by a scalar order-parameter $\omega$ and concomitant
microforces whose working accompanies changes in $\omega$. Since working
is characterized by terms of the form \{ microforce times $\omega^{\hdot}$
\},
microforces are here scalar quantities. Precisely, we assume that
the microforces are described by a (vector) stress \bg{\xi} --whose traction
$\bg{\xi} \cdot {\bf n}$
characterizes forces transmitted across oriented surfaces of unit
normal ${\bf n}$-- in conjunction with a (scalar) body force $\pi$,
which represents
internal forces distributed over the volume of the fluid; these forces
are presumed consistent with the microforce balance \cite{re:fried93}
\begin{equation}
\int_{\partial R} \bg{\xi} \cdot {\bf n} da + \int_{R} \pi dv = 0
\end{equation}
for each control volume $R$, or equivalently
\begin{equation}
\label{eq:xibalance}
{\rm div} \bg{\xi} + \pi = 0.
\end{equation}

A basic hypothesis of the theory is an identification
of the order parameter $\omega$ with the constituent density $\varphi$:
\begin{equation}
\omega = \varphi.
\end{equation}

\subsection{Second law in the form of a dissipation inequality}

We restrict attention to isothermal behavior and
therefore consider a mechanical version of the second law: for
each control volume $R$, the rate at which the energy of $R$ increases
cannot exceed the working on $R$ plus the rate at which energy is
transported to $R$ by diffusion. Let $\mu_{b}$ denote the chemical potential
of constituent $b$, and let
\begin{equation}
\mu = \mu_{1} - \mu_{2}
\end{equation}
Then
\begin{equation}
-\sum_{b=1,2} \int_{\partial R} \mu_{b} {\bf h}_{b} \cdot {\bf n} da =
- \int_{\partial R} \mu {\bf h} \cdot {\bf n} da
\end{equation}
represents energy carried into $R$ across $\partial R$ by diffusion, while
\begin{equation}
\int_{\partial R} {\bf T} ~ {\bf n} \cdot {\bf v} da, ~~~ \int_{\partial R}
\varphi^{\hdot} ~ \bg{\xi} \cdot {\bf n} da
\end{equation}
gives the working of the macroscopic and microscopic stresses. Thus, since
we neglect the relative kinetic energy of the constituents, the appropriate
form of the second law is the dissipation inequality
\begin{equation}
\label{eq:dieq1}
\left\{ \int_{R} ( \psi + k) dv \right\}^{\hdot}  \le \int_{\partial R}
{\bf T} ~ {\bf n} \cdot {\bf v} da + \int_{\partial R} \varphi^{\hdot} ~
\bg{\xi}
\cdot {\bf n} da - \int_{\partial R} \mu ~ {\bf h} \cdot {\bf n} da ,
\end{equation}
where $k = \rho {\bf v}^{2} /2$. The body force $\pi$
does not contribute, since it acts internally to $R$.

By (\ref{eq:pcon}), (\ref{eq:phicon}), (\ref{eq:force}),
and (\ref{eq:xibalance}), (\ref{eq:dieq1}) has the local form
\begin{equation}
\label{eq:dieq2}
\psi^{\hdot} - {\bf T} \cdot \rgrad {\bf v} + (\pi - \mu) \varphi^{\hdot}
+ \bg{\xi} \cdot \rgrad (\varphi^{\hdot}) + {\bf h} \cdot \rgrad \mu \le 0.
\end{equation}
Combining (\ref{eq:dieq2}) with (\ref{eq:mathident}),
(\ref{eq:force}), and (\ref{eq:extras}),
we are led to a local dissipation inequality
\begin{equation}
\label{eq:dieq3}
\psi^{\hdot} - \left[ {\bf S} + ( \rgrad \varphi ) \otimes \bg{\xi} \right]
\cdot \rgrad {\bf v} + (\pi - \mu) \varphi^{\hdot} - \bg{\xi} \cdot
(\rgrad \varphi)^{\hdot} + {\bf h} \cdot \rgrad \mu \le 0
\end{equation}
that will form a basis for our discussion of constitutive equations.

The negative of the left side of
(\ref{eq:dieq3}) represents the dissipation ${\cal D}$,
as its integral over $R$ is
the right side of (\ref{eq:dieq1}) minus the left. Thus, for motion in a
container $B$ with both ${\bf v} = 0$ and  $\varphi^{\hdot} ~  \bg{\xi} \cdot
{\bf n} = 0$ on $\partial B$,
\begin{eqnarray}
\left\{ \int_{B} ( \psi + k) dv \right\}^{\hdot} & = & - \int_{B} {\cal D} dv
\le 0 ~~~ {\rm if} ~~~ {\bf h} \cdot {\bf n} = 0 ~~~ {\rm on } ~~~ \partial
B \nonumber \\
\left\{ \int_{B} ( \psi - \mu_{0} \rho + k) dv \right\}^{\hdot} & = &
- \int_{B} {\cal D} dv \le 0 ~~~ {\rm if} ~~~ \mu = \mu_{0} = {\rm const.}
 ~~~ {\rm on } ~~~ \partial B.
\end{eqnarray}
The thermodynamic development therefore
yields natural Lyapunov functions for certain classes of flows.

\section{Constitutive equations. Restrictions imposed by the Second Law}

In standard theories of diffusion the
chemical potential is given, constitutively, as a function of
the density, but here we consider systems sufficiently far from
equilibrium that a relation of this type is no longer valid;
instead we allow the chemical potential and its gradient to join
the stretching, the density, and the density gradient in the list
of constitutive variables. Precisely, we consider constitutive equations
of the form
\begin{eqnarray}
\label{eq:conlist}
\psi = \hat{\psi}({\bf D}, \varphi, \rgrad \varphi, \mu , \rgrad
\mu), \nonumber \\
{\bf S} = \hat{{\bf S}}({\bf D}, \varphi, \rgrad \varphi, \mu , \rgrad
\mu), \nonumber \\
\bg{\xi} = \hat{\bg{\xi}}
({\bf D}, \varphi, \rgrad \varphi, \mu , \rgrad \mu), \\
\pi = \hat{\pi}({\bf D}, \varphi, \rgrad \varphi, \mu , \rgrad
\mu), \nonumber \\
{\bf h} = \hat{{\bf h}}({\bf D}, \varphi, \rgrad \varphi, \mu ,
\rgrad \mu), \nonumber
\end{eqnarray}
with each of the constitutive functions isotropic, since
the material is a fluid. To avoid notation such as
$\partial_{\rgrad \varphi} \hat{\psi} ({\bf D}, \varphi, \rgrad \varphi, \mu ,
\rgrad \mu)$ for the partial derivative with respect to
$ \rgrad \varphi$, we write
\begin{equation}
{\bf b} = \rgrad \varphi, ~~~ {\bf s} = \rgrad \mu.
\end{equation}
Not all constitutive relations of the form (\ref{eq:conlist})
are admissible, as without further restrictions (\ref{eq:conlist}) will violate
the dissipation inequality (\ref{eq:dieq3}). To determine the requisite
restrictions we choose arbitrary fields ${\bf v}, \varphi$ and $\mu$,
for the velocity, order parameter, and chemical potential, and use
(\ref{eq:conlist})
to compute
a constitutive process consisting of ${\bf v}, \varphi, \mu$ and the fields
$\psi, {\bf S}, \bg{\xi}$, $\pi$ and ${\bf h}$;
such a constitutive process will satisfy (\ref{eq:dieq3})
if and only if
\begin{eqnarray}
\label{eq:dieq4}
\left[ \partial_{{\bf D}} \hat{\psi} ( \ldots ) \right] \cdot {\bf
D}^{\hdot} + \left[ \partial_{\varphi} \hat{\psi} ( \ldots ) + \hat{\pi} (
\ldots ) - \mu \right]  \varphi^{\hdot} + \left[ \partial_{{\bf b}}
\hat{\psi} ( \ldots ) - \hat{\bg{\xi}} ( \ldots ) \right] \cdot
{\bf b}^{\hdot} - \nonumber \\
\left[ \hat{{\bf S}} ( \ldots ) + {\bf b} \otimes \hat{\bg{\xi}} \right]
\cdot \rgrad {\bf v} + \nonumber \\
\left[ \partial_{\mu} \hat{\psi} ( \ldots ) \right]  \mu^{\hdot} + \left[
\partial_{{\bf s}} \hat{\psi} ( \ldots ) \right] \cdot {\bf s}^{\hdot} +
\hat{{\bf h}} ( \ldots ) \cdot {\bf s} \le 0,
\end{eqnarray}
where we have written $( \ldots )$ as
shorthand for the list of independent constitutive variables:
\begin{equation}
( \ldots ) = ({\bf D}, \varphi, \rgrad \varphi, \mu , \rgrad \mu).
\end{equation}
It is possible to find fields ${\bf v}({\bf x},t), \varphi({\bf x},t)$,
and $\mu ({\bf x},t)$ such that ${\bf v}, \rgrad {\bf v},
(\rgrad {\bf v})^{\hdot}, \varphi, \varphi^{\hdot}, {\bf b} = \rgrad
\varphi, {\bf b}^{\hdot} = (\rgrad \varphi )^{\hdot}, \mu, \mu^{\hdot},
{\bf s} = \rgrad \mu$, and ${\bf s}^{\hdot} = (\rgrad \mu )^{\hdot}$
have arbitrarily prescribed values at some
chosen point and time.\footnote{
It is tacit that there are {\em external} mass supplies and forces
available to ensure satisfaction of the balance laws for arbitrary
choices of the fields ${\bf v}, \varphi$ and $\mu$. We chose not to
introduce external fields (source terms) since they tend to complicate
the discussions and since it is only here that they are required.
Specifically, we need an external mass supply $r$ and external body
forces ${\bf b}$ and $\gamma$ such that (\ref{eq:phicon}),
(\ref{eq:force}), and (\ref{eq:xibalance})
become $\varphi^{\hdot} = - \rdiv {\bf h} + r, ~ \rdiv {\bf T} + {\bf b} = \rho
{\bf v}^{\hdot}$,  and $\rdiv \bg{\xi} + \pi + \gamma =0$, and such that
the dissipation inequality (\ref{eq:dieq1})
contains the additional term $\int_{R}
\left\{ {\bf b} \cdot {\bf v} + \varphi^{\hdot} \gamma + \mu r \right\}
dv$ on the right hand side. Allowing an external supply for each balance
law is an assumption now standard in continuum mechanics; such
assumptions are, of course, tacit in derivations dependent upon
arbitrary variations of a field.}
Thus, since the terms ${\bf D}^{\hdot}, \varphi^{\hdot}, {\bf b}^{\hdot},
\mu^{\hdot}$ and ${\bf s}^{\hdot}$ appear linearly in (\ref{eq:dieq4}), it
follows
that $\partial_{{\bf D}} \hat{\psi} = {\bf 0},
\partial_{\varphi} \hat{\psi} = \mu - \hat{\pi},
\partial_{\mu} \hat{\psi} = 0, \partial_{{\bf s}} \hat{\psi}
= {\bf 0}$, and $\partial_{{\bf b}} \hat{\psi} = \hat{\bg{\xi}}$.
We are therefore led to
the following constitutive restrictions:
\begin{itemize}
\item[(i)] the
free energy and microstress are independent of ${\bf D}, \mu$, and $\rgrad
\mu$,
and are related through
\begin{equation}
\label{eq:xidef}
\hat{\bg{\xi}}(\varphi, \rgrad \varphi) = \partial_{{\bf b}} \hat{\psi}
(\varphi, \rgrad \varphi);
\end{equation}
\item[(ii)]
the internal microforce is independent of ${\bf D}$ and $\rgrad \mu$ and
represents a nonequilibrium contribution to the chemical
potential:
\begin{equation}
\label{eq:pidef}
\hat{\pi}(\varphi, \rgrad \varphi, \mu) = \mu - \partial_{\varphi}
\hat{\psi} (\varphi, \rgrad \varphi);
\end{equation}
\item[(iii)] the following inequality must be satisfied for all
values of the arguments:
\begin{equation}
\label{eq:dieq5}
\hat{{\bf h}} ( \ldots ) \cdot \rgrad \mu - \left[ \hat{{\bf S}} ( \ldots ) +
{\bf b} \otimes \hat{\bg{\xi}} ( \varphi, \rgrad \varphi) \right] \cdot
\rgrad {\bf v} \le 0.
\end{equation}
\end{itemize}

Since the skew part of $\rgrad {\bf v}$ can be chosen arbitrarily and
independently of ${\bf D}$, and since ${\bf S}$ is symmetric, $\bg{\xi} =
\hat{\bg{\xi}} (\varphi, {\bf b})$  satisfies
\begin{equation}
{\bf b} \otimes \bg{\xi} = \bg{\xi} \otimes {\bf b}, ~~~ ({\bf b} \cdot
\bg{\xi} ) {\bf b} = | {\bf b} |^{2} \bg{\xi}
\end{equation}
and is hence parallel to ${\bf b}$. Further, the isotropy of
$\bg{\xi} = \hat{\bg{\xi}}(\varphi,{\bf b})$  implies
that $\bg{\xi} = {\bf 0}$ when ${\bf b}= {\bf 0}$; hence there is a scalar
function $\alpha (\varphi, {\bf b})$  such that
\begin{equation}
\label{eq:xib}
\bg{\xi} = \alpha (\varphi, {\bf b} ) {\bf b}.
\end{equation}
Let
\begin{equation}
{\bf P} = \hat{{\bf P}}( \ldots ) = \hat{{\bf S}} ( \ldots ) +
\alpha (\varphi, {\bf b}) \left[ {\bf b} \otimes {\bf b} -
\frac{1}{3} | {\bf b} |^{2} {\bf 1} \right] ;
\end{equation}
then, since ${\rm tr} {\bf D} =0$, (\ref{eq:dieq5}) reduces to
\begin{equation}
\label{eq:dieq6}
\hat{{\bf h}} ( \ldots ) \cdot {\bf s} - \hat{{\bf P}} ( \ldots ) \cdot
{\bf D} \le 0.
\end{equation}
${\bf P}$ in (\ref{eq:dieq6}), represents a
thermodynamic stress, as it is conjugate to the stretching ${\bf D}$.

Any set of constitutive equations consistent with the restrictions
(\ref{eq:xidef}), (\ref{eq:pidef}), (\ref{eq:xib}), and (\ref{eq:dieq6})
will be consistent with the dissipation
inequality (\ref{eq:dieq3}). Our purpose here is not to develop the most
general theory possible, but rather to develop a theory that couples the
essential features of the Cahn-Hilliard and Navier-Stokes equations.
Newtonian fluids have stress linear in ${\bf D}$, while Cahn-Hilliard
diffusion has mass flux linear in $ \rgrad \mu$ and free energy quadratic
in $\rgrad \varphi$. Guided by these theories and by
(\ref{eq:dieq6}), we now assume that
the thermodynamic stress ${\bf P}$, and the mass flux ${\bf h}$ have the
specific forms
\begin{equation}
{\bf P} = 2 \nu (\varphi ) {\bf D}, ~~~ {\bf h} = - m (\varphi ) \rgrad \mu,
\end{equation}
with mobility $m( \varphi )$ and viscosity $\nu ( \varphi )$ nonnegative,
and, appealing to (\ref{eq:xib}), that the free energy $\psi$ has the form
\begin{equation}
\psi = f(\varphi) + \frac{1}{2} \alpha (\varphi) | \rgrad \varphi |^{2}
\end{equation}
with $\alpha ( \varphi )$ nonnegative. Then
\begin{eqnarray}
{\bf T} & = & - p {\bf 1} + 2 \nu (\varphi) {\bf D} - \alpha (\varphi) \rgrad
\varphi \otimes \rgrad \varphi, \nonumber \\
\bg{\xi} & = & \alpha ( \varphi ) \rgrad \varphi, \\
\pi & = & \mu - f'(\varphi) - \frac{1}{2} \alpha^{\prime}(\varphi)
| \rgrad \varphi |^{2} \nonumber,
\end{eqnarray}
where, for convenience, the pressure $p$ has been replaced by
$p + (\alpha /3) | \rgrad \varphi |^{2}$.
If we assume, in addition, that $\alpha, m$, and $\nu$ are constants, then
\begin{eqnarray}
\label{eq:cons}
\psi & = & f(\varphi) + \frac{1}{2} \alpha | \rgrad \varphi |^{2} \nonumber \\
{\bf T} & = & - p {\bf 1} + 2 \nu {\bf D} - \alpha  ~ \rgrad \varphi \otimes
\rgrad \varphi \nonumber \\
\bg{\xi} & = & \alpha  ~ \rgrad \varphi, \\
\pi & = & \mu - f^{\prime} (\varphi) \nonumber \\
{\bf h} & = & -m ~ \rgrad \mu.
\end{eqnarray}
The term $\alpha ~ \rgrad \varphi \otimes \rgrad \varphi$ in the stress gives
rise to normal stresses in the absence of flow; this term should
represent surface tension and could effect boundary conditions at
a free surface in a nonstandard manner.

Further, and what is most important, the microbalance (\ref{eq:xibalance})
and the constitutive relation
for $\pi$ yield Cahn's formula for the chemical potential
\begin{equation}
\label{eq:mu}
\mu = f' (\varphi) - \alpha \Delta \varphi.
\end{equation}

The basic PDEs of
the theory follow upon substituting (\ref{eq:cons}) and
(\ref{eq:mu}) into (\ref{eq:phicon})
and (\ref{eq:force}); the results, after replacing the original pressure
$p$ by $p - (\alpha /6) | \rgrad \varphi |^{2}$, consist of
generalized Navier-Stokes equations
\begin{equation}
\label{eq:final}
\rho {\bf v}^{\hdot} = - \rgrad p + \nu \Delta {\bf v} - \alpha
( \Delta \varphi) \rgrad \varphi, ~~~ \rdiv {\bf v} = 0,
\end{equation}
coupled to the Cahn-Hilliard equation
\begin{equation}
\label{eq:ch2}
\varphi^{\hdot} = m \Delta \left[ f' (\varphi) - \alpha \Delta \varphi
\right].
\end{equation}
The latter equation depends explicitly on the flow velocity, since
the material time derivatives have spatial forms:
\begin{equation}
\varphi^{\hdot} = \varphi_{t} + ( \rgrad \varphi) \cdot {\bf v}, ~~~
{\bf v}^{\hdot} = {\bf v}_{t} + (\rgrad {\bf v}) {\bf v}.
\end{equation}

An interesting alternative
form for these PDEs mentions the chemical potential explicitly:
\begin{eqnarray}
\rho {\bf v}^{\hdot} & = & - \rgrad p + \nu \Delta {\bf v} + \mu ~ \rgrad
\varphi, ~~~ \rdiv {\bf v} = 0, \nonumber \\
\varphi^{\hdot} & = & m \Delta \mu \\
\mu & = & f^{\prime} ( \varphi ) - \alpha \Delta \varphi , \nonumber
\end{eqnarray}
where the original pressure $p$ has now been replaced by
$p-( \alpha /6) | \rgrad \varphi |^{2} + f(\varphi)$.

\section{Immiscible fluids described by an order parameter}

With but minor modifications the theory
presented above can be applied to immiscible fluids. Here the order
parameter $\varphi$  has a given value in each fluid, with the fluid interface
defined by the variation in $\varphi$ between values. Since the fluids are
immiscible, the interface should be transported with the fluid; we
therefore require that
\begin{equation}
\label{eq:phiim}
\varphi^{\hdot} = 0,
\end{equation}
so that $\varphi$ is constant on streamlines. Further, we let
$\rho = \rho(\varphi)$,
allowing the fluids to have different densities; this dependence
on $\varphi$ is consistent with incompressibility, since, by
(\ref{eq:phiim}), $\rho (\varphi)^{\hdot} = 0.$

The microforce balance is no longer relevant; the basic
equations are the classical balances (\ref{eq:pcon}) and the dissipation
inequality
\begin{equation}
\left\{ \int_{R} \left( \psi + k \right) dv \right\}^{\hdot} \le
\int_{\partial R} {\bf T} {\bf n} \cdot {\bf v} da ~~~
k = \frac{1}{2} \rho(\varphi) {\bf v}^{2},
\end{equation}
and this leads to the local dissipation inequality
\begin{equation}
\label{eq:dieq7}
\psi^{\hdot} - {\bf S} \cdot \rgrad {\bf v} \le 0.
\end{equation}

As constitutive equations we assume that
\begin{equation}
\psi = \hat{\psi} ( {\bf D}, \varphi,\rgrad \varphi ), ~~~
{\bf S} = \hat{{\bf S}}({\bf D}, \varphi, \rgrad \varphi),
\end{equation}
and proceeding as before, consistency with (\ref{eq:dieq7}) leads, by virtue of
(\ref{eq:mathident}) and (\ref{eq:phiim}), to the inequality
\begin{equation}
\left[ \partial_{{\bf D}} \hat{\psi}({\bf D},\varphi, {\bf b}) \right]
\cdot {\bf D}^{\hdot} - \left[ \hat{{\bf S}}({\bf D},\varphi, {\bf b}) +
{\bf b} \otimes \partial_{{\bf b}} \hat{\psi}
({\bf D},\varphi, {\bf b}) \right] \cdot \rgrad {\bf v} \le 0,
\end{equation}
rendering the free energy
independent of {\bf D}. Finally, restricting attention to energies that
are independent of $\varphi$ and quadratic in $\rgrad \varphi$, we are led to
the constitutive equations
\begin{eqnarray}
\psi & = & \frac{1}{2} \alpha | \rgrad \varphi |^{2}, \nonumber \\
{\bf T} & = & - p {\bf 1} + 2 \nu (\varphi) {\bf D} -
\alpha \left\{ \rgrad \varphi \otimes
\rgrad \varphi - \frac{1}{3} | \rgrad \varphi |^{2} 1 \right\} .
\end{eqnarray}
(A dependence of $\psi$ on $\varphi$ does not alter the resulting
PDEs.)

The basic equations, after replacing $p$ by
$p - (\alpha /6) | \rgrad \varphi |^{2}$, therefore have the form
\begin{eqnarray}
\label{eq:immiscible}
\rho (\varphi) {\bf v}^{\hdot} & = & - \rgrad p + \nu (\varphi) \Delta
{\bf v} + 2 \nu^{\prime}(\varphi) {\bf D} \rgrad \varphi -
\alpha ( \Delta \varphi) \rgrad \varphi, \nonumber \\
\rdiv {\bf v} & = & 0, ~~~ \varphi^{\hdot} = 0.
\end{eqnarray}
These equations may be useful in the study of
the interface between immiscible fluids; the order parameter
automatically tracks the interface, here a layer, and provides
a capillarity term $\alpha ( \Delta \varphi ) \rgrad \varphi$
that should model surface tension. However, if the initial condition
for $\varphi$ involves an interface of finite width (or, for practical
purposes, in any numerical scheme), there is no
mechanism in (\ref{eq:immiscible}) to avoid spreading of the layer.
Hence frequent renormalizations of the layer would be required in an actual
computation.

\section{A numerical example: coarsening}

As a numerical application of the theory, we consider
the kinetics of coarsening following a quench into the coexistence
region of a binary fluid in two space-dimensions (see ref.
\cite{re:mullins89} for a review of
coarsening). Numerical solutions of the Cahn-Hilliard equation {\it without}
hydrodynamic effects have been obtained
by a number of authors, for both critical \cite{re:gawlinski89} and
off-critical \cite{re:toral89} quenches, while hydrodynamic effects on
spinodal decomposition have been discussed in
ref. \cite{re:koga91}. We
consider here the off-critical region of the phase diagram in which the
decay of the original state involves the formation of localized droplets.

Our starting point is Eqs. (\ref{eq:final}) and (\ref{eq:ch2}) with,
\begin{equation}
f \left( \varphi \right) = -\frac{r}{2} \varphi^{2} + \frac{u}{4} \varphi^{4},
\end{equation}
as the homogeneous part of the free energy density, with $r$ and $u$
positive constants.
For incompressible flows in two dimensions, it is convenient to introduce
a stream function $\zeta$, such that, $
{\bf v} = \frac{\partial \zeta}{\partial y} {\bf i} - \frac{\partial
\zeta}{\partial x} {\bf j}$,
where ${\bf i}$ and ${\bf j}$, respectively,
are the unit vectors in the $x$ and $y$ directions.
We next scale the order parameter by $\sqrt{r/u}$, lengths by the the mean
field correlation length $\xi = \sqrt{\alpha /r}$, and time by the diffusion
time $\tau = \alpha / mr^{2}$. In what follows, all variables are
assumed to be dimensionless. Taking the curl of (\ref{eq:final}),
and restricting
attention to small Reynolds number ($Re = mr / \nu $),
so that the inertial term  in (\ref{eq:final}) is negligible, we
obtain,
\begin{equation}
\label{eq:biharmonic}
\Delta^{2} \zeta + C \left[ \rgrad \left( \Delta \varphi \right) \times
\rgrad( \varphi ) \right] = 0,
\end{equation}
where $C = 3 \sigma \xi / 2 \eta mr ,~ (\eta = \rho \nu) $ plays
the role of a capillary number.
$\sigma = 2 \alpha r / 3 \xi u $ is the mean field value of the
surface tension in this model. The fluid is enclosed in a
square cavity of side $L$, with the fluid velocity satisfying no-slip boundary
conditions on the walls of the cavity. Hence, the boundary conditions
appropriate for (\ref{eq:biharmonic}) are that both $\zeta $ and
its normal derivative $\partial \zeta / \partial n$ vanish on the boundary.

The dimensionless Cahn-Hilliard equation reads,
\begin{equation}
\label{eq:ch}
\partial_{t} \varphi + {\bf v} \cdot \rgrad \varphi =  -
\Delta \left[ \varphi - \varphi^{3} + \Delta \varphi \right],
\end{equation}
and is supplemented by the boundary conditions
$\partial \varphi / \partial n = \partial \mu / \partial n$ = 0
where $\mu = - [ \varphi - \varphi^{3} + \Delta \varphi ]$.

The equations (\ref{eq:biharmonic}) and (\ref{eq:ch}) are solved
numerically in a square, two dimensional grid with a backward implicit method,
and a fast biharmonic solver (further
details are given in  \cite{re:chella94}).
We take $C = 10$, and use the same parameters for the simulation
as those of \cite{re:toral89} for a solution of the Cahn-Hilliard
equation at an off-critical value of the order parameter:
$L = 256$, a grid spacing $\delta x = 1.0$, and an initial configuration
in which $\varphi$ is random, gaussianly distributed with mean $\left<
\varphi \right> = 1/ \sqrt{3}$ and variance unity. We have chosen a variable
integration step $\delta t = 0.01$ for $0 \leq t < 4.0$, $\delta t =
0.1$ for $4.0 < t \leq 664$ and $\delta t = 0.5$ for $664 < t \leq
3700$.

Our results are summarized in Figures \ref{fi:drops} -- \ref{fi:growth}.
At short times, drops of the minority phase are seen to nucleate, grow, and
coarsen so as to decrease the free energy of the system. Figure
\ref{fi:drops} shows the distribution of phases --as defined by the
dimensionless order parameter-- as a function of time, while Figure
\ref{fi:stream} displays the corresponding distribution for the stream
function. The main qualitative difference between our results and those
for the uncoupled Cahn-Hilliard equation is the flow-induced coalescence
of neighboring droplets. The times in Figures \ref{fi:drops} and
\ref{fi:stream}
are chosen to illustrate coalescence. Figure \ref{fi:stream} shows large a
localized variations of the stream function in the regions in which
coalescence is occurring. Coalescence is not observed in the purely
relaxational Cahn-Hilliard equations. There coarsening is dominated by
diffusion.

Figure \ref{fi:growth} graphs the total droplet-perimeter, per unit area,
with the droplet boundaries defined as the level set $\varphi = 0$. At
sufficiently long times growth is faster than the classical
Lifshitz-Slyozov law (growth law of the form $t^{1/3}$) for coarsening
in the absence of hydrodynamic effects. Siggia \cite{re:siggia79}
predicted that hydrodynamic effects would affect coarsening, and San
Miguel, Grant and Gunton \cite{re:sanmiguel85} reexamined Siggia's
argument for the case of two dimensions. They argued that the dominant
coarsening mechanism in three dimensions (which is expected to lead to a
growth law in which the average size of the domains is proportional to time)
is not operative in two dimensions. Droplet coalescence was suggested to
be dominant instead, and on phenomenological grounds
they deduced a $t^{1/2}$ growth law; Figure \ref{fi:growth} is in accord with
this result. Our results are also in agreement
with recent molecular dynamics simulations of
coarsening in binary fluids \cite{re:leptoukh95}, but
disagree with a previous numerical study of the same model in two space
dimensions, and inside the unstable region of the phase diagram, that
obtained a growth exponent of 0.69 \cite{re:farrell89}.

\section*{Acknowledgments}
The work of Gurtin and Polignone was supported by the Army Research
Office and by the National Science Foundation. A portion of the research
of Gurtin was carried out during a visit to the Institute for
Advanced Study. The work of Vi\~nals was supported by the Microgravity
Science and Applications Division of the NASA, and in part by the
Supercomputer Computations Research
Institute, which is partially funded by the U.S. Department of Energy.

\begin{figure}
\caption{Order parameter values (in grey scale) as a function of time
showing nucleation and coarsening of the structure. Times have been
chosen to highlight coalescence events.}
\label{fi:drops}
\end{figure}

\begin{figure}
\caption{Stream function $\zeta$ in grey scale at the same times shown
in Fig. {\protect \ref{fi:drops}}. For the values of the parameters used,
$\zeta$ lies approximately between -0.5 and 0.5. The regions of largest
variation in the stream function are located in the regions where coalescence
takes place.}
\label{fi:stream}
\end{figure}

\begin{figure}
\caption{Inverse perimeter-density as a function of time for $C = 10$. The
three solid lines indicate power laws with the value of the exponent
shown.}
\label{fi:growth}
\end{figure}

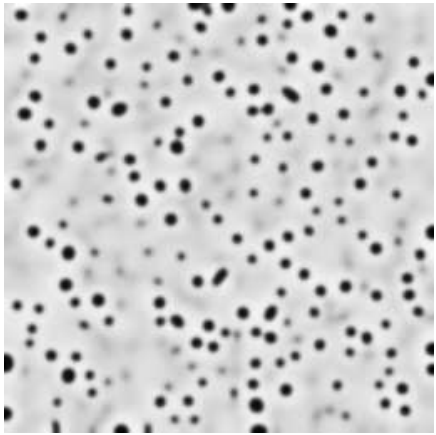
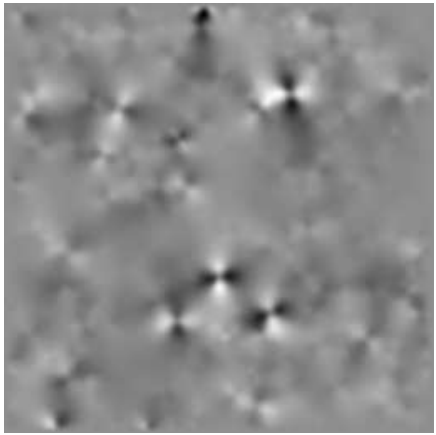
\end{document}